# The fluid dynamics of liquid mushrooms

**Akshay Manoj Bhaskaran[1], Arnov Paul[1], Apurba Roy[1], Devranjan Samanta[2] and Purbarun Dhar[1, *]**

[1]Hydrodynamics and Thermal Multiphysics Lab (HTML), Department of Mechanical Engineering, Indian Institute of Technology Kharagpur, West Bengal – 721302, India

[2]Department of Mechanical Engineering, Indian Institute of Technology Ropar, Punjab – 140001, India

* *Corresponding author*: E–mail: *purbarun@mech.iitkgp.ac.in*

## Abstract

Droplets that impact the surface of a deep liquid pool may form a vertical jet after the cavity formation event, provided they have sufficient impact energy. Depending on the associated time scales and the effect of the Rayleigh-Plateau instability, this jet may either continue to rise, or may form satellite droplets via necking. Collision of these structures with a second incoming droplet, ejected from the same dispensing tip as the first droplet, may result in the formation of various lamellar patterns, depending on the impact conditions – giving rise to liquid mushroom and/or umbrella structures. In this research, we experiment with hydrodynamics of such liquid mushrooms, and study the effect of droplet impact height, surface tension, and viscosity on the dynamics of such lamellar formations. We further explore the  role of the orientation of incoming droplet impact, i.e. whether head-on or offset collision with the rising jet/satellite droplet. We discuss the spatio-temporal evolution of the lamella diameters, and its susceptibility to surface tension, viscosity, and droplet impact height. We put forward a theoretical model based on energetics, to predict the maximum spread diameter of the lamellae, which yields accurate predictions with respect to our experiments. Our findings may help to provide important insights towards a fluid dynamic phenomenon observed often in nature and may be important in niche utilities as well.

## 1.  Introduction

The events associated with droplet impact, and its consequences, is an important fundamental phenomenon in nature, and for utilities of practical and industrial significance. Interactions of a droplet or a family of droplets with substrates has been studied extensively owing to its diverse applications;  such as cleaning up oil spills [1], spray cooling in electronics and in metallurgy [2,3], inkjet printing [4], and fuel injection in internal combustion engines [5], in environment-friendly crop spraying, [], tuning phase-change characterisitcs, [], to name a few. This observation is also relevant in welding applications, i.e., deposition of molten metal droplet on the surface of the weld pool [6].

All such droplet impact studies can be classified in accordance to the nature of the target surface; such as on a dry, solid surface [7], on thin liquid films, [8] and on deep liquid pools [9].In the case of deep liquid pools, the focus of the studies has been on specific regimes; such as splashing, crown formation [10], cavity evolution [11], bubble entrainment, Rayleigh jet formation, pinch-off and subsequent satellite droplet formation [12]. Research on the impact of single droplet, irrespective of the target, has been done primarily for normal impact, with subsequent studies for variation in viscosity, surface tension, and height. Some works have also considered cases where droplet impact is oblique in nature [13, ], which is observed in practical situations such as rain impact on windshields, spray cooling, etc.

Typically, these droplet interactions exhibit complex dynamics, which have often fascinated artists, photography enthusiasts, popular science writers, and the community of fluid dynamics researchers across the globe. On droplet ejection from very close to a liquid-air interface, such that it has low kinetic energy, partial coalescence of the droplet occurs, followed by a cascading coalescence and fragmentation into subsequently smaller drops [14]. Taking this a step further, studies were done where the droplets were released from higher heights. The consequent higher impact energy resulted in the formation of a Rayleigh jet from the surface of the pool, and subsequently broke-up into satellite droplets by the Rayleigh-Plateau (RP) instability [15]. Studies have revealed that droplet impact on solid geometrical targets display a direct correlation between the shape of the target and the filaments formed via rim instability []; and it depends on the relative strength between the perturbations of the imposed target and the highest unstable mode of RP instability [16].

While works have mostly focussed on stationary pools, studies have also classified regimes associated with the impact of droplets on moving liquids [17]. Deviating from the use of identical liquid for pool and droplet to understand the specific structures inherent to the liquid, works have used immiscible liquids to study the nature on splashing [18]. Reports have analysed the jet dynamics and associated energetics, further describing the jet velocity using Froude number and capillary velocity [19]. Studies using energy balance have described the jet formation in terms of energy conservation; using cavity depth, jet height, and impact velocity of droplet; and  concluding that the majority of the impact energy is lost via dissipation, and only ~28 % is utilized in cavity formation, which is then directly transferred to the jet [20]. Energy based models have also been used to predict the maximum spread of droplets on cylindrical and spherical targets, and good accuracy was noted [21, 22]. Hence energetics based event prediction has been a successful approach in droplet impact research.

Thus far, experimental studies have focussed on the impact dynamics on a pool, up till the regime of satellite droplet generation by rising jet pinch-off, by virtue of the RP instability [15, 23]. But beyond this, neither the interaction of a second impacting droplet interacting with the rising jet nor with the satellite droplet has been studied. When a jet rises from a pool (due to a droplet impact), and is impacted upon by a second incoming droplet, liquid mushrooms or umbrellas may form. It is plausible that the very niche applications of such impacts, accentuated by the fact that perfect repetations of such events require extensive trial experiments, could be a reason for the absence of such studies in literature. These impact structures are to be avoided in many practical applications, such as in droplet-on-demand thermal management of electronic hot spots, where the rebounding droplet may collide against an incoming droplet, resulting in insufficient cooling [24], and also in welding processes, where similar interactions could result in weaker weld patches due to non-uniform deposition of the molten metal droplets[25].

In this research, we experiment with, and study the formation and fluid dynamics of such liquid mushroom/umbrella structures. We explore the effects of varying the release heights of the droplet ($h$), the surface tension ($\sigma$), and viscosity ($\mu$) of the liquid (we have used the same liquid for the droplets and the pool). We vary the surface tension by the addition of a surfactant, Sodium Dodecyl Sulphate (SDS), to water. We have considered SDS concentrations ranging from 0 to 1 times the critical micelle concentration (CMC) (refer table 1 for surface tension values [26]). Similarly, a broad range of viscosity of the fluid was obtained by adding different concentrations of pure Glycerol to water (refer table 1 for viscosity values [27,28]). In our study we have emphasized on two initial regimes: (i) head-on impact, where the second incoming droplet falls along the central axis of the rising jet (or satellite drop), and (ii) offset impact, where the second droplet falls offset from the central axis. We qualitatively and quantitatively study the fluid dynamics of the events, and put forward a theoretical model to predict the maximum spread state of the liquid mushrooms for different impact scenarios.

## 2. Materials and methods

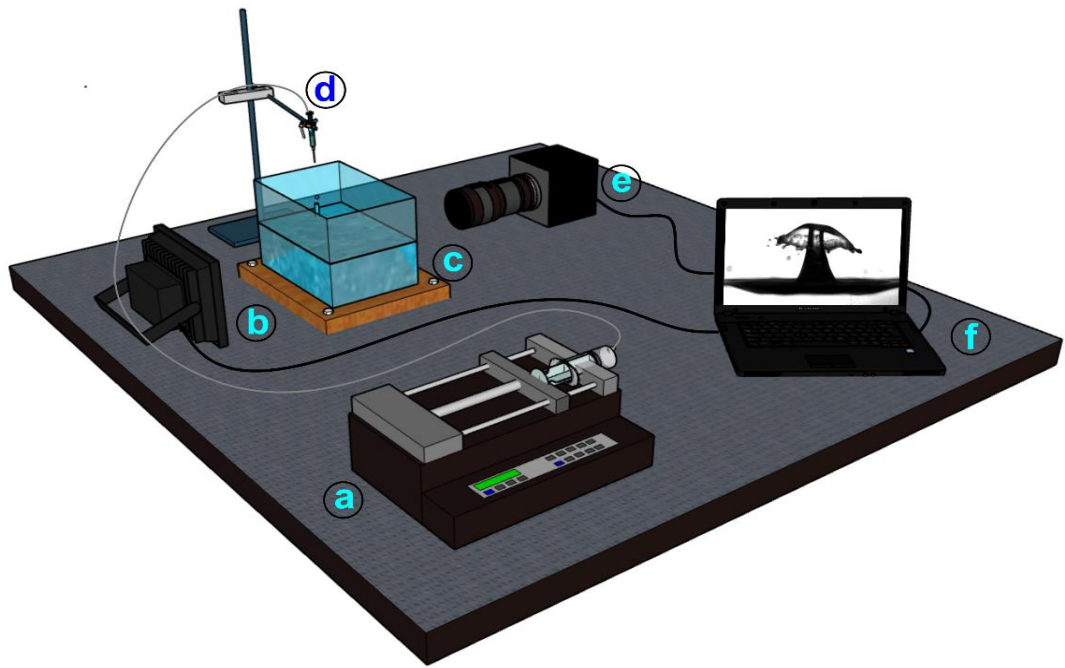

**Fig. 1:** Schematic of the experimental setup: (a) syringe pump, (b) backlight, (c) liquid pool, (d) dispenser system, (e) high-speed camera, (f) computer. The entire assembly is housed on a vibration-free table.

Fig. 1 shows the schematic of the experimental setup. A cubical container made of highly transparent, thin acrylic sheets, was used to hold the pool. The free-surface area and height of the pool was large enough to eliminate any wall effects (each dimension of the pool was >30 times the droplet diameter). Droplets were generated using a precision syringe pump (Harvard Apparatus) at varying flow-rates, as per the required time gaps and regime (i.e., rising jet or satellite impact), using a flexible tubing and dispensing needle with a diameter of ~3 mm. A high speed camera (Photron Inc.) connected to a 105 mm fixed focal length macro lens captured the events at 8000 frames per second. The backlight illumination was provided by a 50 W lamp with a diffuser sheet. The camera was set to capture the phenomena happening at and above the pool interface (lateral view). All the experiments were carried out at room temperature and the setup was mounted on a vibration-free table.

**Table 1:** Properties of test liquids (all aqueous solutions) at room temperature. The label is used to represent the liquid sample in the text of the article.

| Test liquid | $D_{01}$ (mm) | $D_{02}$ (mm) | $\mu$ (mPa.s) | $\rho$ (kg/m$^3$) | $\sigma$ (mN/m) | Label |
|---|---|---|---|---|---|---|
| Water | 4.764 | 4.447 | 0.89 | 1000 | 72 | Water |
| SDS 0.075CMC | 4.443 | 4.335 | 0.89 | 1000 | 66 | Water 0.92X ST |
| SDS 0.15CMC | 4.312 | 3.935 | 0.90 | 1000 | 63 | Water 0.88X ST |
| SDS 0.25CMC | 4.579 | 4.048 | 0.90 | 1000 | 57 | Water 0.8X ST |
| SDS 0.50CMC | 4.120 | 3.724 | 0.92 | 1000 | 46 | Water 0.6X ST |
| SDS 1CMC | 2.777 | 2.271 | 0.93 | 1000 | 33 | Water 0.5X ST |
| Glycerol 47% | 4.105 | 3.409 | 4.47 | ~1120 | 72 | Water 5X V |
| Glycerol 59% | 4.345 | 3.942 | 8.93 | ~1150 | 70 | Water 10X V |
| Glycerol 69% | 4.953 | 4.884 | 17.86 | ~1170 | 67 | Water 20X V |

Due to the inherent dynamic nature of these complicated impact events, the observations would require fine-tuning and hence several runs for each experiment were conducted to extract data of significance. The properties of the different liquids, along with the first and second droplet diameter, $D_{01}$ and $D_{02}$ respectively, are presented in Table 1.Although we vary the flow rates, for a given liquid, the size of the droplets ($D_{01}$ or $D_{02}$) remained almost constant (variations $< 5\%$) as the dispenser needle remained the same. We deduce the impact velocity of the droplet as $U = \sqrt{2g(h - h_{ji})}$, where $h$ is the droplet ejection height, and $h_{ji}$ indicates the jet height (determined from image processing). We have varied the release height, $h$, from 5 cm to 35 cm (in steps of 5 cm). To generate offset impacts, the needle was minutely tilted during the experiments. Our experiments generated four different outcomes: (1) head-on collision with rising jet, (2) head-on collision with satellite droplet, (3) off-set collision with rising jet, and (4) off-set collision with satellite droplet.

In Table 1 the right-most column provides the labels used for the various test liquids. Here, for liquids names in the format "Water AX ST", A $= \frac{\sigma_{liquid}}{\sigma_{water}}$, X implies the multiplier, and ST denotes surface tension. Hence, Water 0.88X ST signifies a liquid whose surface tension is 0.88 times that of water. Likewise, for liquids labelled "Water BX V", B $= \frac{\mu_{liquid}}{\mu_{water}}$, and V refers to viscosity. Hence, Water 10X V represents a liquid with viscosity 10 times that of water. As earlier studies have reported on the impact dynamics of droplets on liquid pools, the focus of this work would purely be on the collision dynamics of the second droplet with the rising jet/satellite drop, and not on events occuring before that. We henceforth refer to the second droplet diameter, $D_{02}$ as $D_d$ to simplify the representation. The time evolution of the events is analysed using the non-dimensional time, $\tau = \frac{tU}{D_d}$, where t is the instantaneous time. The other associated dimensionless terms used are the Weber number $\left( We_d = \frac{\rho U^2 D_d}{\sigma} \right)$, Reynolds number $\left( Re_d = \frac{\rho U D_d}{\mu} \right)$, and Capillary number $\left( Ca_d = \frac{\mu U}{\sigma} \right)$.

## 3. Results and discussions

### 3.1 General observations:

In order to represent the dynamics, we made arrays corresponding to the time evolution of various lamellar structures after a collision of the second-incoming droplet with jet/satellite drop. This was done to make specific emphasis on the effects of droplet ejection height, surface tension, and viscosity.

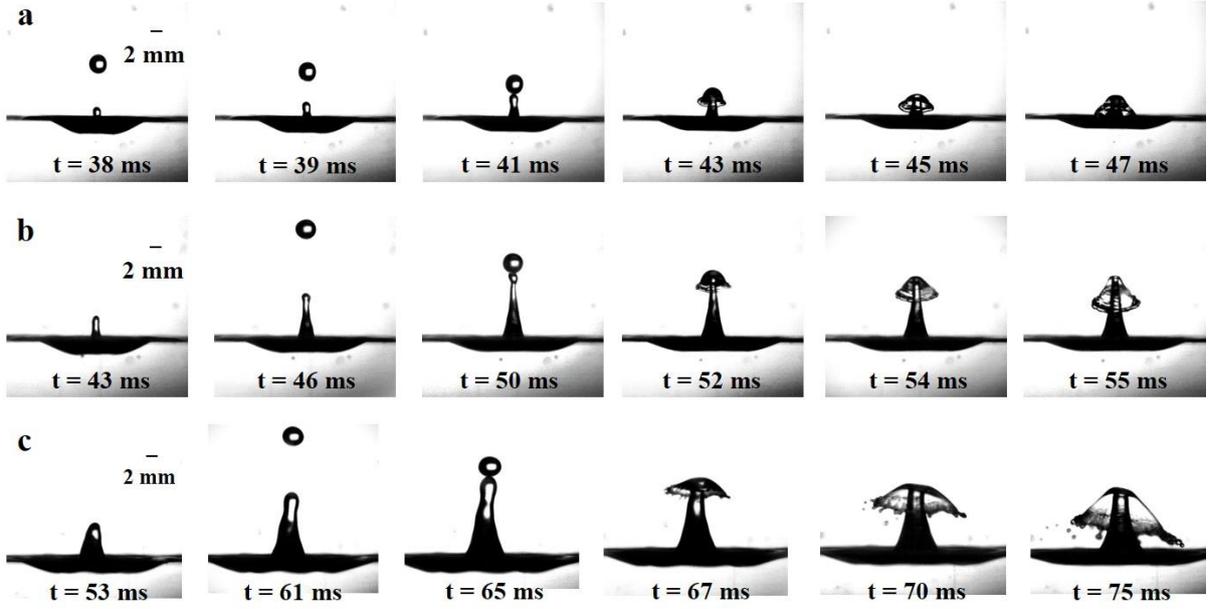

**Fig. 2:** Mushroom/umbrella formation dynamics for varying droplet ejection heights, for head-on collision with jet: (a) h=10 cm, (b) h=20 cm and (c) h=30 cm.

Fig. 2 to Fig. 5 represents the effect of droplet release height $h$, on the formation of hydrodynamic structures corresponding to water. Fig. 2 represents the head-on collision of the second droplet with the jet. As the height increases from 10 cm to 30 cm, the jet height as well as the splash diameter of the umbrella increases. This is due to the elevated kinetic energy possessed by the initial as well as the second droplet, as $U \sim \sqrt{h}$. The impact energy of the first droplet (kinetic energy + surface energy) is utilized in the formation of the cavity (~28%, remaining lost via dissipation [30]). This energy possessed by the cavity is directly transferred to the jet formation process. The impact energy of the second droplet after collision reflects in the form of inertia, which contributes to the enhanced spreading of the umbrella. It can be observed from Fig. 2a to Fig. 2c that the maximum umbrella diameter ($D_u$) increases significantly with increasing ejection height. Another fascinating observation is the formation of filaments at higher heights (Fig. 2c). This is because, after the collision, the liquid due to inertia moves radially outwards, such that a liquid rim is formed around the edge [16]. Now, if the ejection height is higher, the corresponding inertia on the droplet after collision would be higher. This would imply more liquid would spread to this rim. As a result, the area of the rim increases. Now this enhanced umbrella area would correspond to an increase in surface energy. Subsequently, the surface tension of the liquid would try to reduce this surface energy by breaking this region into smaller filaments, as observed in Fig. 2c. Therefore, the radial spreading of these umbrella structures are due to the combined effects of inertia, surface tension, and viscous forces.

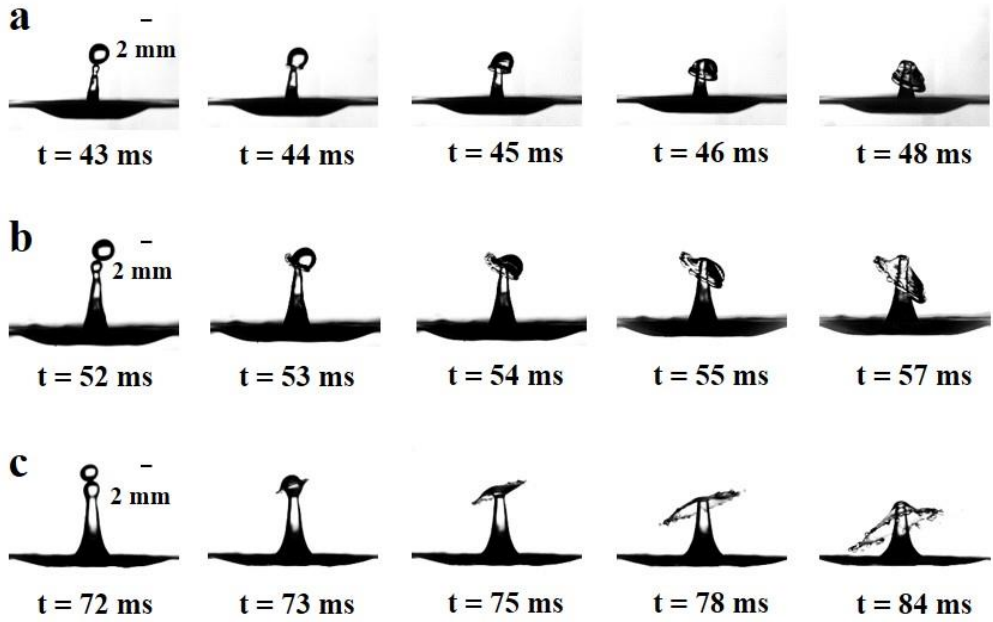

**Fig. 3:** Mushroom/umbrella formation events for varying droplet ejection heights for offset collision with rising jet: (a) 10 cm, (b) 20 cm, (c) 30 cm.

Fig. 3 highlights the impact dynamics with the same kinematic conditions except the second droplet impacts the rising jet in an off-set regime. The effect of off-set becomes more prominent for higher release heights. At low release heights, as the droplet diameter is comparable to the jet height, there is no notable distinction between Fig. 3a from Fig. 2a. In Fig. 3b and 3c, we notice a truncation in the umbrella diameter at the top extreme compared to the bottom extreme. This asymmetry can be understood by observing Fig. 3c at t = 73 ms. The second droplet gets sliced into two hemispheres, where the top portion is in contact with air on one end and is seemingly sliding against the interface with bottom half. The bottom hemisphere on the other hand, is in contact with the jet, and experiences higher viscous dissipation compared to the top hemisphere (spread area is higher). This results in truncation of the bottom hemisphere which then forms the top extreme of the umbrella.

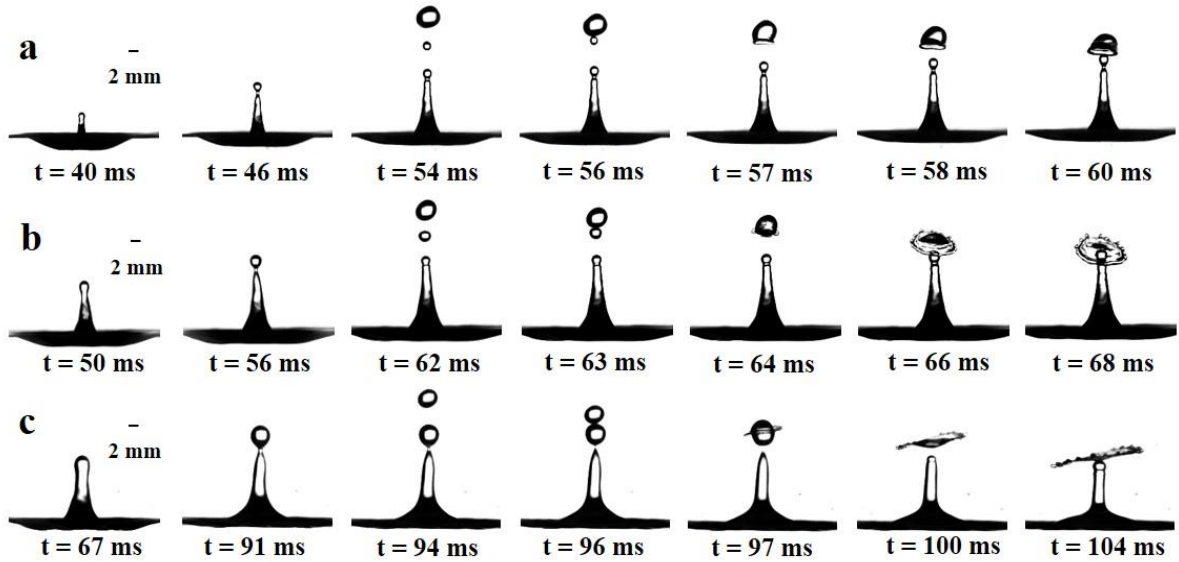

**Fig. 4:** Lamellae formation events for varying droplet ejection heights, for head-on collision with the satellite droplet: (a) h=10 cm, (b) h=20 cm and (c) 30 cm.

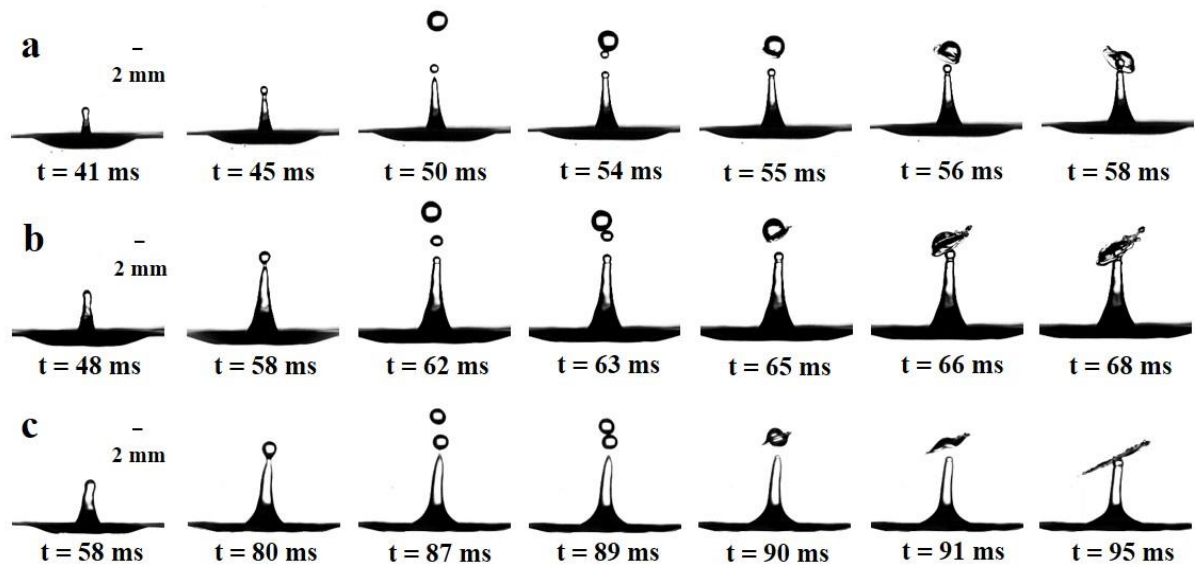

**Fig. 5:** Lamellae formation events for varying droplet ejection heights, for offset collision with the satellite droplet: (a) h=10 cm, (b)h= 20 cm and (c) h=30 cm.

Fig. 4 and Fig. 5, depict the satellite droplet collision cases corresponding to Fig. 2 and Fig. 3, respectively. To generate satellite droplets at the same heights as rising jets, we have to set appropriate flow rates. Here, the flow rate of the liquid should be lower than that used in the rising jet cases. At a lower flow rate, it takes more time for the liquid to flow into the dispenser and increase the size (hence, weight) of the droplet. As a result, this allows sufficient time for the jet to undergo necking via Rayleigh-Plateau instability. This difference in curvature (radial and axial) on the jet creates a region of high pressure surrounding the neck. Now as to whether pinch-off of necked region occurs depends on a lot of factors. From previous works, we know that when the wavelength of fluctuations is long enough, pinch-off of smaller droplets occurs [15]. This pinched-off droplet is referred to as the satellite droplet.

This droplet then collides with the delayed second droplet to form lamellar structures (resembles a hat at smaller heights as observed in Fig. 4a, Fig. 4b, Fig. 5a, and Fig.5b, while it takes on a disc shape at higher heights (Fig. 4c and Fig. 5c). We observe that the splash diameter increases with ejection height. The rest of the dynamics are quite similar to the collision with a rising jet.

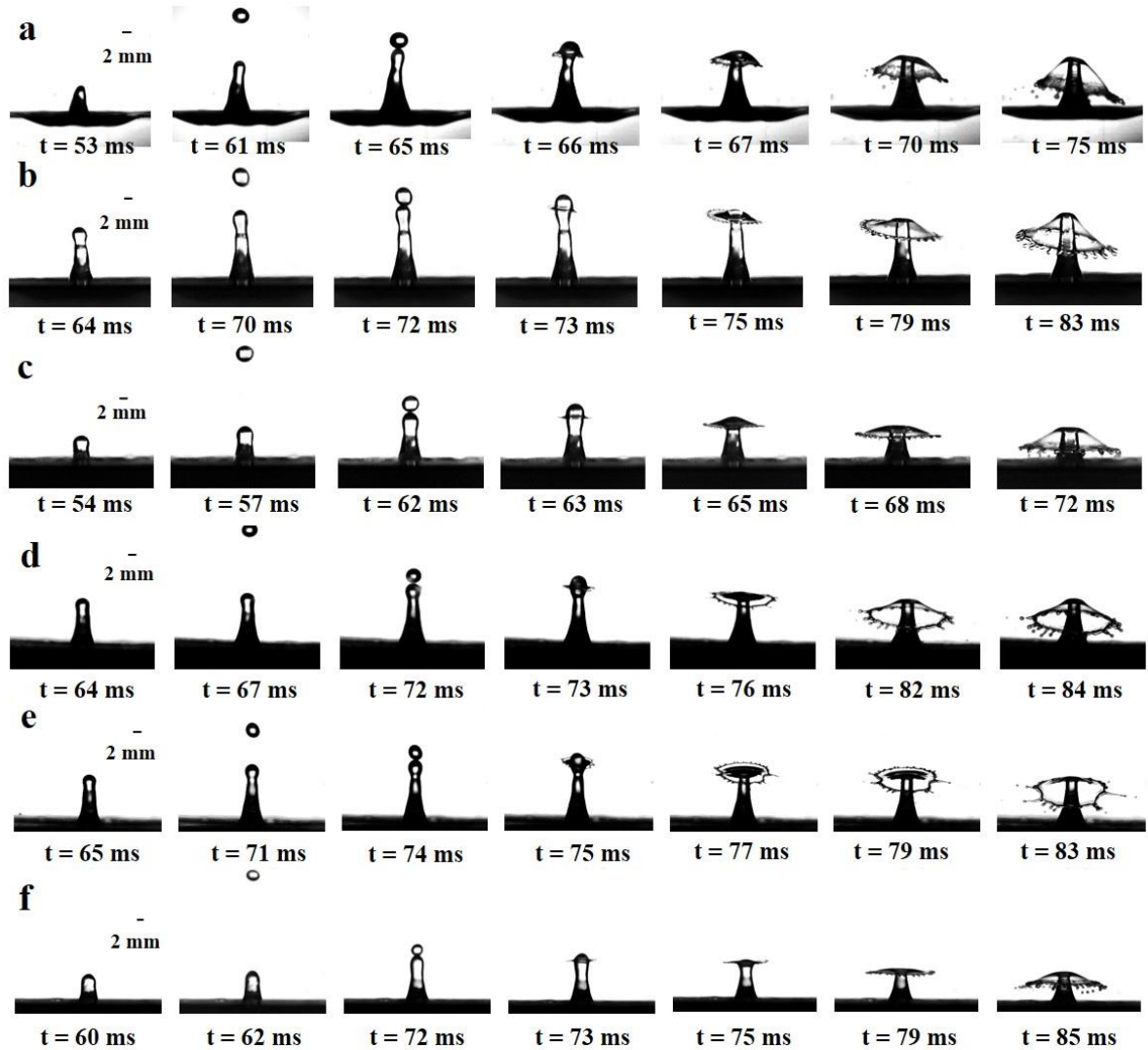

**Fig. 6:** Time evolution of second droplet head-on collision with the rising jet at $h = 30$ cm, for varying surface tension: (a) Water, (b) Water 0.92X ST (c) Water 0.88X ST, (d) Water 0.8X ST, (e) Water 0.6X ST, (f) Water 0.5X ST. Surface tension decreases from top to the bottom row.

Fig. 6, represents the effect of surface tension on the time evolution of umbrella structures formed via second droplet collision with a rising jet, at $h = 30$ cm. There are a few observations that can be made, starting with droplet diameter. While concurrently referring to Table. 1 and Fig. 6, we observe that as surface tension reduces, the equilibrium droplet

diameter decreases due to the reduction in capillary length scale $\sqrt{\frac{\sigma}{\rho g}}$. This initially sounds counter-intuitive as surface tension tries to reduce surface area, but here we observe that the droplet diameter is smallest for Water 0.5X ST. This can be understood from the droplet ejection process. The droplet is released when the weight of the droplet balances the surface tension force experienced along the circumference of the dispenser. As surface tension reduces (Water to Water 0.5X ST), the weight of the droplet required decreases.

As a results, droplets of smaller diameter are ejected for fluids of lower surface tension. The reduced size of the droplets results in lesser kinetic energy as well as surface energy. Therefore, overall impact energy is reduced. We know that the energy of the jet is directly correlated to this impact energy. This energy reduction is reflected by the lower jet heights from Fig. 6a to Fig. 6f.

Another interesting observation would be the filament formation. We know the surface tension tries to oppose the spread of the umbrella and surface tension is responsible for the breakup of the rim into filaments, which may further disintegrate into smaller droplets via Rayleigh-Plateau instability [15].

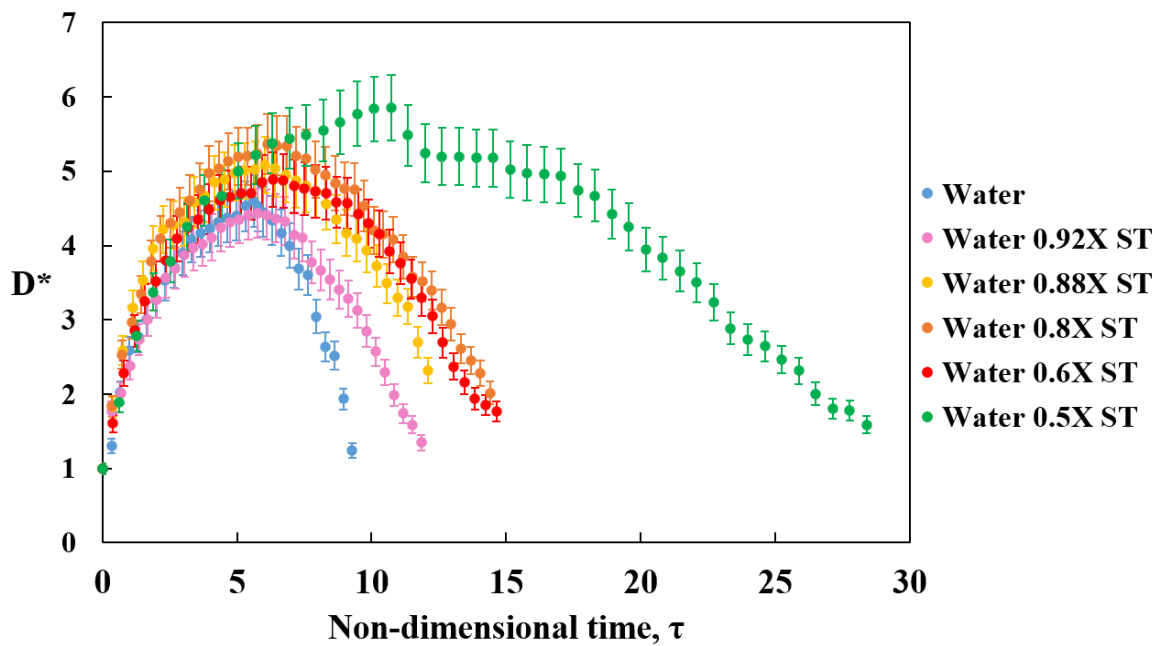

**Fig. 7:** Variation of normalized splash diameter $D^*$ with $\tau$ at $h = 30$ cm, for varying surface tension: head-on collision of second droplet with rising jet.

Although from Fig. 6, the specific effect of surface tension of splash diameter is not very clear from the depicted umbrellas, Fig. 7 helps to get a better understanding. Fig. 7, plots the time evolution of the second droplet after a head-on collision with a rising jet for various values of surface tension, at $h = 30$ cm. To study the dynamics of this time evolution we define a non-dimensional term, referred to as the normalized splash diameter, $D^* = \frac{D_i}{D_d}$; where $D_i$ is an intermediate umbrella diameter at any instant from $\tau(=\frac{tU}{D_d})=0$ till the umbrella eventually collapses onto the jet. Here, $\tau=0$, corresponds to the instance when the second ejected droplet makes first contact with the rising jet. Note that $D_d$ varies with the surface tension of the liquid, which necessitates defining normalized terms like $D^*$ and $\tau$. We observe that for all surface tension values, as time progresses, the intermediate splash diameter $D_i$ increases till it achieves a maximum umbrella diameter, $D_u$ (which corresponds to $D^*{}_{max}\left(=\frac{D_u}{D_d}\right)$ on Fig. 7), beyond which it falls again. We observe that $D^*{}_{max}$ increases with a decrease in surface tension. This is because lower surface tension corresponds to reduced cohesive forces between the liquid molecules. As a result, the liquid molecules are less weakly bound to each other. So, the molecules are splattered to greater extents by inertial forces yielding higher splash diameters. We also observe that as $\sigma$ decreases, the time taken for $D^*$ to reduce from $D^*{}_{max}$ and collapse onto the jet increases. This is because, in the higher surface tension umbrellas, the liquid molecules trying to spread out are pulled back to a greater extent by the bulk liquid molecules. So, they spread out less and deflate quicker.

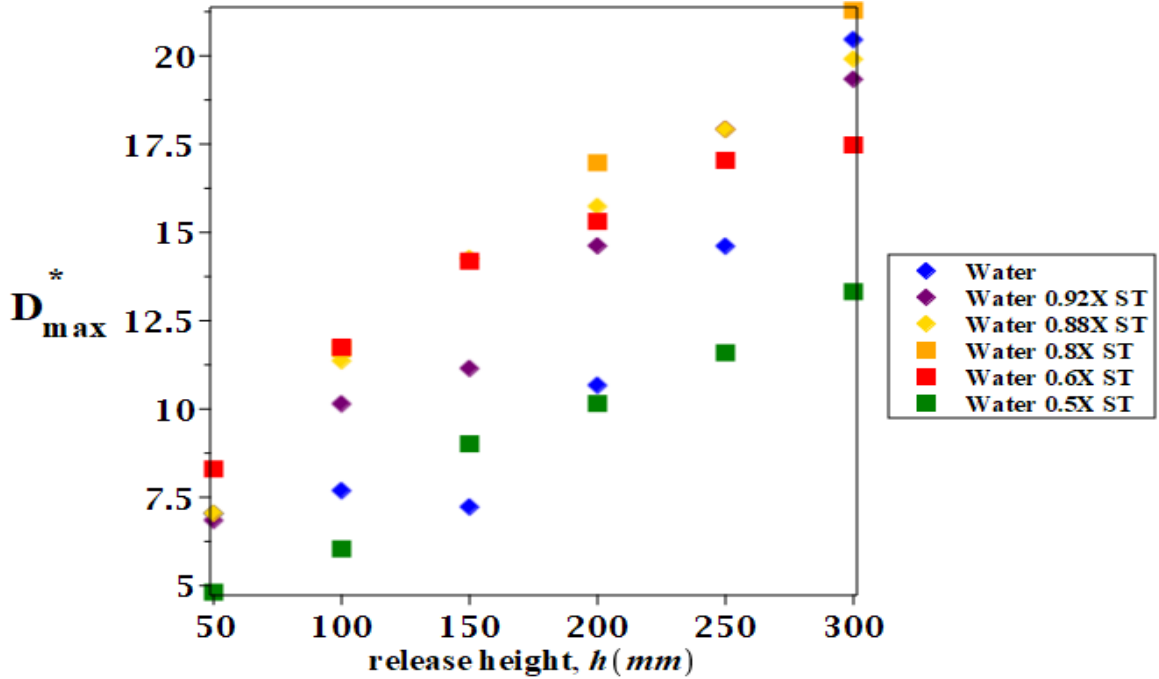

**Fig. 8:** Variation of maximum normalized splash diameter $D^*{}_{max}$ with ejection height, $h$ for varying surface tension: a head-on collision of the second droplet with rising jet.

Fig. 8 represents the variation of $D^*{}_{max}$, with droplet release height for a head-on collision of rising jet with a second droplet. At first glance, Fig. 7 and Fig. 8, seem to contradict as we see that despite the low cohesion and increase in $U$ (i.e. increase in $h$), the lowest normalized splash diameter is observed for the lowest surface tension case (i.e. Water 0.5X ST). On the other hand, from water to water 0.6X ST, we observe $D^*{}_{max}$ to increase with release height, as expected. The anomalous behaviour of water 0.5X ST, is owing to the small size of the ejected droplet. As the volume of liquid present itself is low, the extent of spreading correspondingly decreases. In Fig. 7, we observed it to have the highest spreading primarily because the evolution time was also scaled against $\frac{D_d}{U}$. This time scale is much smaller for water 0.5X ST compared to the other surface tension cases, due to small $D_d$. Further, we observe that at higher ejection heights, the splashing becomes more inertia driven, i.e. the enhanced mass of the droplet overpowers surface tension effects. Lower surface tension droplets have smaller sizes and hence lower inertial energy.. Therefore, despite the reduced cohesive forces, since mass is lower, energy is lower.

Fig. 9 represents the effect of viscosity on the time evolution of umbrella structures formed via second droplet collision with rising jet, at $h = 30$ cm. One prime observation is that as viscosity increases from Fig. 9a to 9b, there is no formation of filaments. The enhanced viscous forces suppresses the filament formation from the rim.

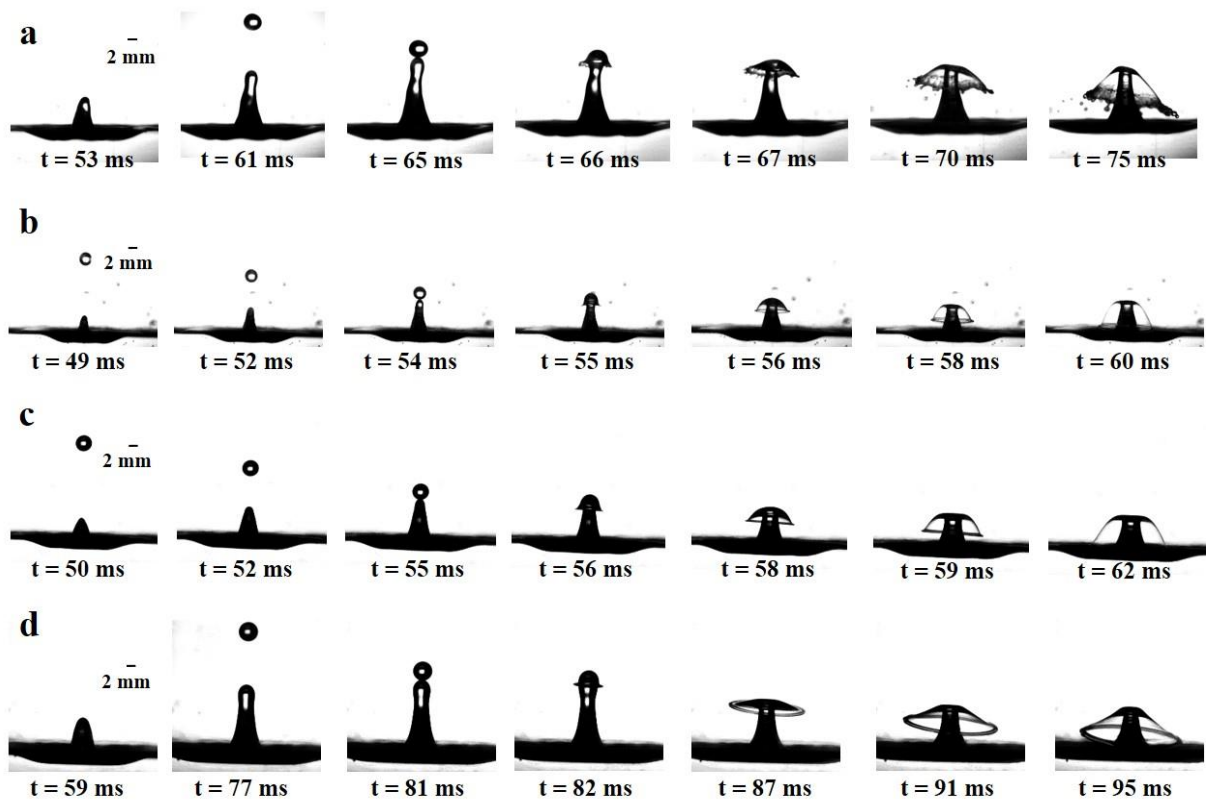

**Fig. 9:** Time evolution of second droplet collision with the rising jet at $h = 30$ cm, for varying viscosity: (a) Water, (b) Water 5X V, (c) Water 10X V, (d) Water 20X V.

Based on the experimental results presented in Fig. 9, it is readily evident that the splash diameter of the umbrella increases with an increase in viscosity. Quantitative estimates of the $D^*$ for varying viscosity at $h = 30$ cm in fig.10 further emphasizes this observation. This seems counter-intuitive as the viscous forces would oppose the growth of umbrella splash diameter. A plausible explanation could be associated with air entrapment by the high viscous umbrella. With the increase in viscosity, the liquid behaves as an amorphous solid. So, as the second incoming droplet approaches the rising jet, entrapment of the air sheet might take place between them [31, 32]. As a result, the area over which viscous dissipation occurs is effectively reduced. This later is found to be the case while solving for the energetics of high viscous fluids in section 3.2. Hence, as viscosity increases the splash diameter is found to increase.

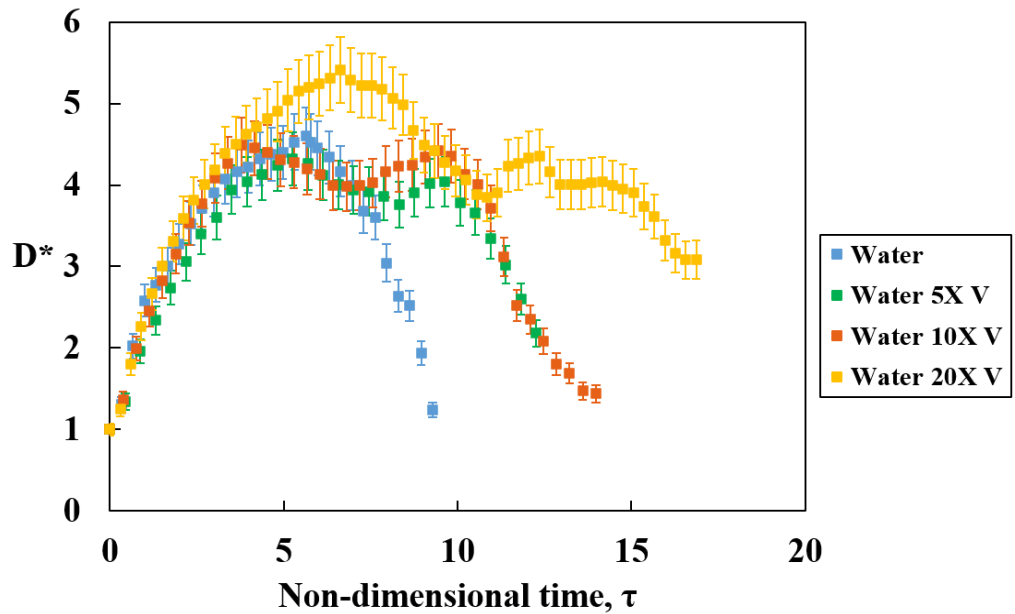

**Fig. 10:** Variation of normalized splash diameter $D^*$ with $\tau$ at $h = 30$ cm, for varying viscosity.

Fig. 11 describes the same situation but for varying droplet ejection heights. We observe that as the ejection height increases, the maximum normalized splash diameter increases. Further, for all the heights, we observe that with an increase in viscosity, $D^*_{max}$ increases. At higher heights, more inertial energy as well as a greater extent of air entrapment results in higher splash diameter.

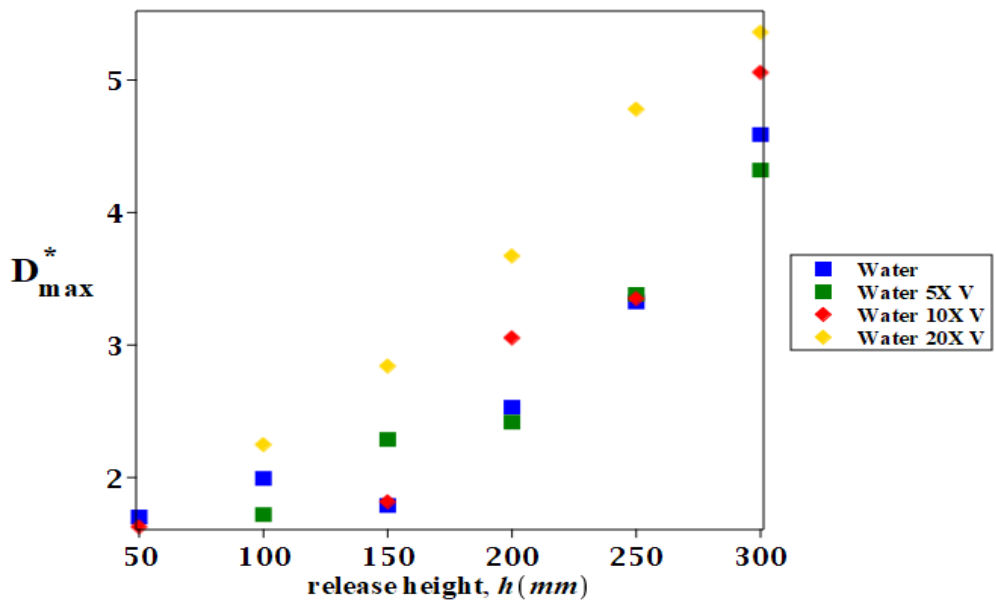

**Fig. 11:** Variation of maximum normalized splash diameter $D^*_{max}$ with ejection height, $h$ for varying viscosity.

In Fig. 12, the umbrella structures of various test liquids (Water, Water 0.6X ST, and Water 10X V) are compared against each other for head-on and off-set collisions at $h = 30$ cm. A clear transition is observed from Fig. 12a to Fig. 12f. The lower surface tension case i.e. Water 0.6X ST (Fig. 12c and 12d) and water (Fig. 12a and 12b) display the formation of filaments and subsequent breakup of the filaments into smaller droplets. While, the high viscosity case (Fig. 12e and Fig. 12f) represents an umbrella without any ejection of filaments from its rim.

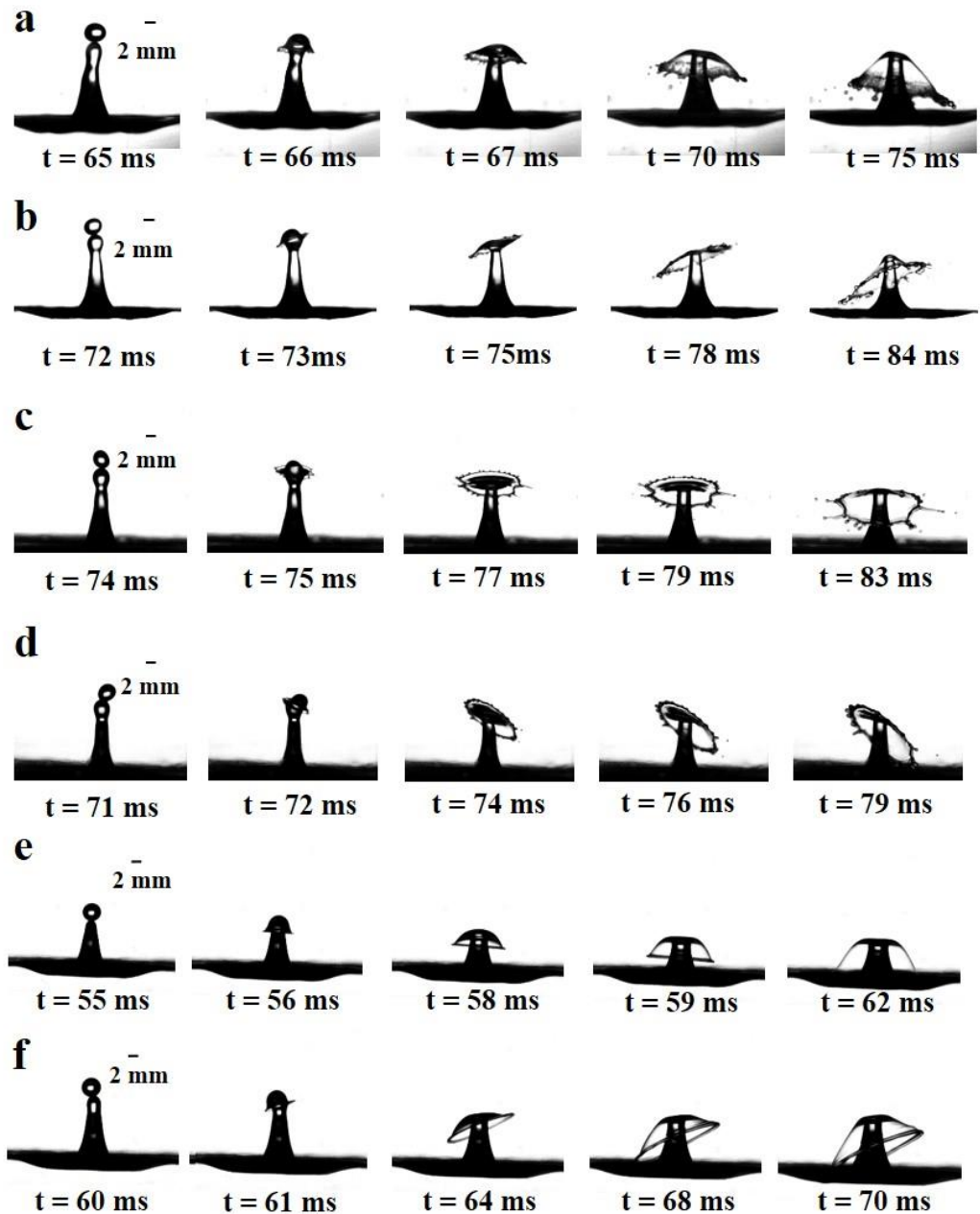

**Fig. 12:** Comparison of lamellar structure formation between head-on and off-set collisions with jet, at $h = 30$ cm: (a) Water (head-on), (b) Water (off-set), (c) Water 0.6X ST (head-on), (d) Water 0.6X ST (off-set), (e) Water 10X V (head-on), (f) Water 10X V (off-set).

Fig. 13 and Fig. 14 represent the time evolution of a second droplet collision with a satellite drop, for surface tension and viscosity variation, respectively. Similar to the previous discussion, the surface tension results are characterized by the formation of filaments along the rim, while viscosity cases have no such tendril formation. It could also be observed that the path of a satellite droplet, at each time instant, can be projected onto a cycloidal curve.

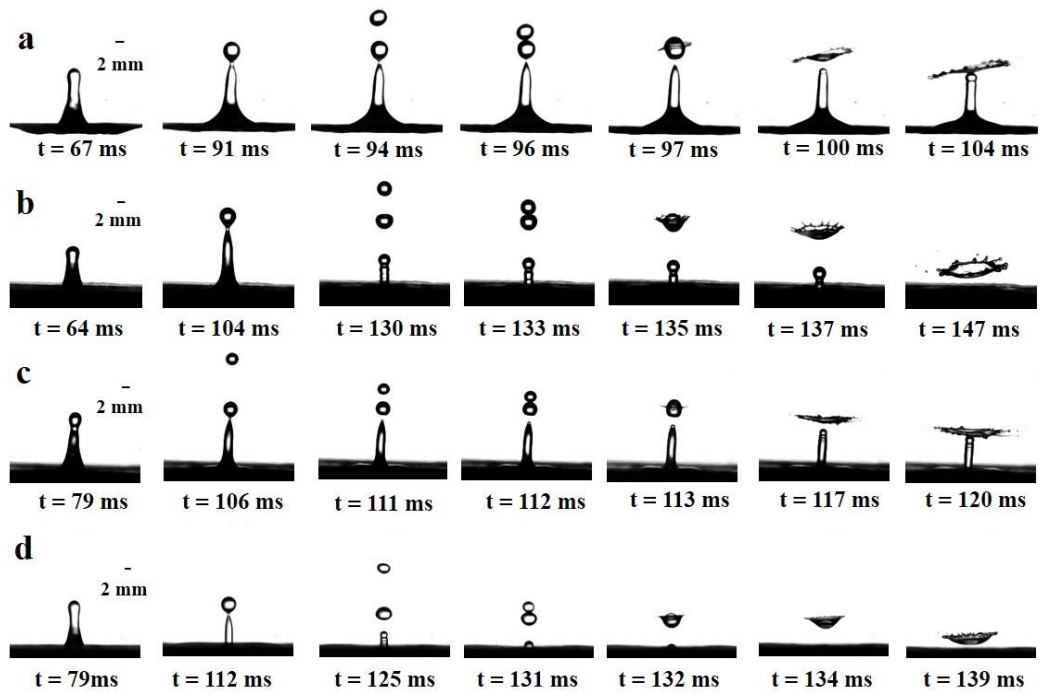

**Fig. 13:** Time evolution of second droplet collision with satellite drop at $h = 30$ cm, for varying surface tension: (a) Water, (b) Water 0.8X ST, (c) Water 0.6X ST, (d) Water 0.5X ST.

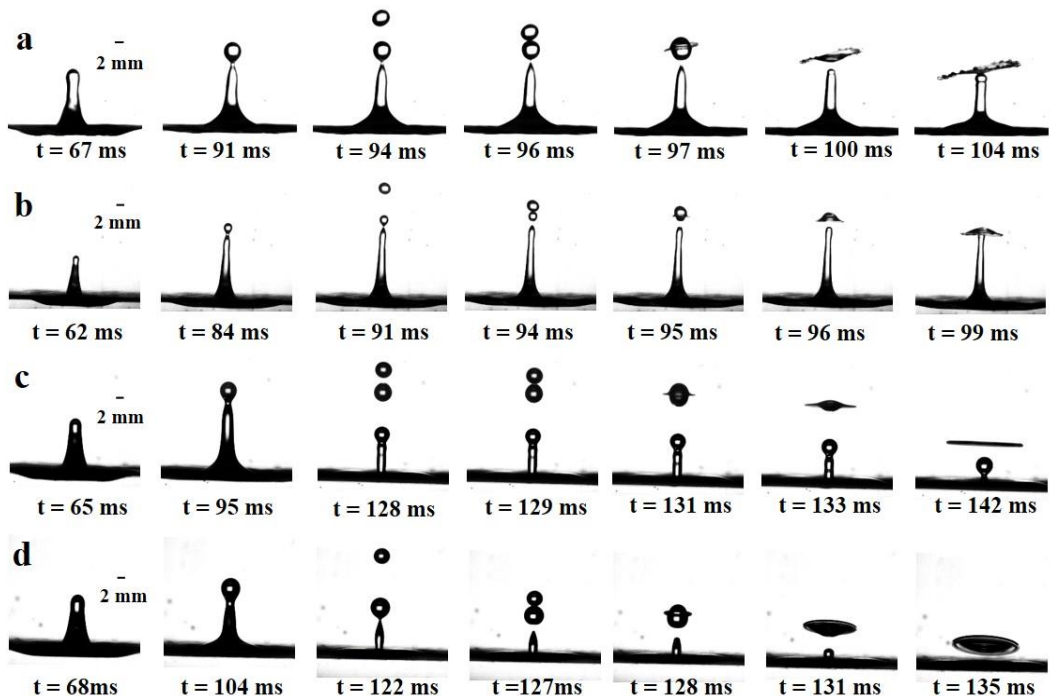

**Fig. 14:** Time evolution of second droplet collision with satellite drop at $h = 30$ cm, for varying viscosity: (a) Water, (b) Water 5X V, (c) Water 10X V, (d) Water 20X V.

## 3.2 Energetics of the umbrella

Subsequently we have developed an analytical model based on energy conservation to predictthe maximum spreading of the droplets after impact. We have focused on the second droplet collision, and made the corresponding measurements required to make the estimate using ImageJ. Further, the collision is depicted using the schematic in Fig. 15, which would be used as the reference for our derivations.

As seen in Fig. 15a, the second droplet of diameter $D_d$ impacts the rising jet of height $h_{j1}$ and diameter $D_j$ with velocity $U$. As a result of the collision, a portion of the jet coalesces with this droplet and spreads to form an umbrella-like lamellar structure, as seen in Fig. 15b. The jet height has now been reduced to $h_{j2}$, and the lamellar structure spreads from $D_d$ to a maximum diameter $D_u$.

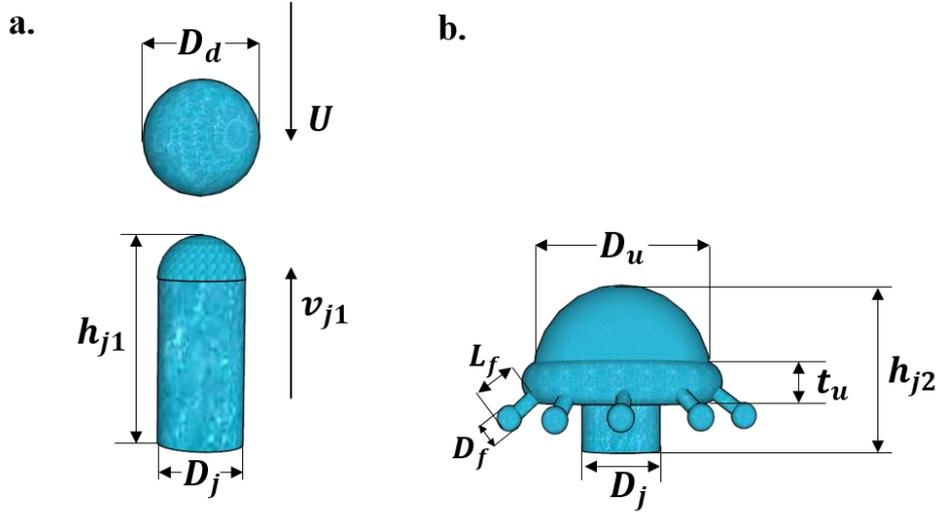

**Fig. 15:** Schematic representing the maximum splash diameter in droplet-jet collision: (a) Before impact, (b) After collision.

For this, the energy prior to collision is given by

$$k \cdot e_d \; + \; s \cdot e_d \; + \; k \cdot e_{j1} \; + \; s \cdot e_{j1} \tag{1}$$

Where, $k \cdot e_d$ ($= \frac{1}{2}\rho V_d U^2$) is the kinetic energy of the second ejected droplet; $s \cdot e_d$ ($= A_d \sigma$) is the surface energy of the droplet; $k \cdot e_{j1}$ ($= \frac{1}{2}\rho V_{j1} v_{j1}{}^2$) is the kinetic energy of the rising jet; and $s \cdot e_{j1}$ ($= A_{j1}\sigma$) is the surface energy associated with the jet. Therefore Eqn. (1) can be expanded as

$$\frac{1}{2}\rho V_d U^2 + A_d \sigma + \frac{1}{2}\rho V_{j1} v_{j1}^2 + A_{j1}\sigma \tag{2}$$

Energy post-collision is given by

$$s \cdot e_u + s \cdot e_{j2} + k \cdot e_{j2} + v_{diss} \tag{3}$$

Where, $s \cdot e_u \ (= (2A_u - A_b + A_{rim} + A_f - A_{fb})\sigma)$ is the total surface energy of the formed umbrella; $s \cdot e_{j2} (= A_{j2}\sigma)$ is the surface energy of the remaining non-coalesced portion of the jet; $k \cdot e_{j2} (= \frac{1}{2}\rho V_{j2} v_{j2}^2)$ is the kinetic energy of the remaining portion of the jet; and $v_{diss}$ is the viscous dissipation work done by the droplet on spreading against the jet.

$$(2A_u - A_b + A_{rim} + A_f - A_{fb})\sigma + A_{j2}\sigma + \frac{1}{2}\rho V_{j2} v_{j2}^2 + v_{diss} \tag{4}$$

Here, the potential energy contribution was found to be negligible compared to the other energy terms for Eqn. (1) and Eqn. (3). Further, a previous study by Mitra et al [33] has shown that it is indeed small during spreading over millimetre sized targets .

Before proceeding with the energy balance, we describe each term in Eqn. (2) and Eqn. (4), in order to simplify the procedure.

The kinetic energy of the second droplet is affected by the height reached by the jet prior to collision and is given by

$$k \cdot e_d = \frac{1}{2}\rho V_d U^2 = \rho \left(\frac{\pi D_d^3}{6}\right) g(H - h_{j1}) \tag{5}$$

Where, $U = \sqrt{2g(H - h_{j1})}$ is the impact velocity, and $H$ is droplet ejection height.

The surface energy of the droplet is given by

$$s \cdot e_d = A_d \sigma = \pi D_d^2 \sigma \tag{6}$$

Now, the jet is formed as a result of the impact of initial droplet with the liquid-pool. Of this around ~ 28% is utilized in the formation of the cavity, which is directly translated into the energy of the jet [30]. As a result, the jet of volume, $V_{j1}$ moves upwards with velocity, $v_{j1}$. The potential energy of the jet is extremely small and hence neglected. The corresponding kinetic energy of the jet is therefore given as

$$k \cdot e_{j1} = \frac{1}{2}\rho V_{j1} v_{j1}^2 = \frac{\rho}{2}\pi \left(\frac{D_j^2}{4}\right) h_{j1} v_{j1}^2 \tag{7}$$

The corresponding surface energy of the jet is given by

$$s \cdot e_{j1} = A_{j1}\sigma = (\pi D_j h_{j1} + \frac{\pi}{4}D_j^2)\sigma \tag{8}$$

Now after collision, the lamellar structure has various components that contribute to surface energy, which includes the area of the umbrella, which is approximated to be a semi-ellipsoid with the base as a circle. This is given as $A_u = \pi D_u h_u$, where $h_u$ is the umbrella height and $D_u$ is the maximum umbrella diameter. $A_b = \pi\left(\frac{D_j^2}{4}\right)$ is the base area of the jet-apex and is being subtracted as this area is in contact with the jet, and we know surface tension between two interfaces of same fluid is zero. Hence there would be no contribution to surface energy. Now the umbrella, as seen in Fig. 15b, assumes a certain film thickness which is given by $t_u = (V_d + V_{neck})/A_u$. As mentioned earlier, when the droplet collides with the jet, a portion of the jet also coalesces with the incoming droplet to spread and form the lamellar structure. Now, due to this coalescence, the height of the jet reduces from $h_{j1}$ to $h_{j2}$. Therefore, this volume that undergoes coalescence is given by $V_{neck} = \pi\left(\frac{D_j^2}{4}\right)(h_{j1} - h_{j2})$. Therefore, the area of the rim formed, as seen in Fig. 15b, is given by $A_{rim} = \pi D_u t_u$. Now depending on the surface tension in play, there is a possibility of formation of filaments along the liquid rim. The total area of the filaments (assumed as cylindrical) is given by $A_f = n\left(\pi D_f L_f + \pi\left(\frac{D_f^2}{4}\right)\right)$, where $n$ is the number of filaments, $D_f$ is the average diameter of each filament, $L_f$ is the average length of each filament. Now, an additional area $A_{fb} = n\pi\left(\frac{D_f^2}{4}\right)$, has to be subtracted from the rim as this region is in contact with filaments of the same liquid; hence won't contribute to surface energy.

Using all this, the surface energy of the umbrella can be represented as

$$s \cdot e_u = \left(2A_u - A_b + A_{rim} + A_f - A_{fb}\right)\sigma = $$
$$\left( \begin{array}{c} 2\pi D_u h_u - \pi\left(\frac{D_j^2}{4}\right) + \left(\frac{\frac{\pi D_d^3}{6}+\pi\left(\frac{D_j^2}{4}\right)(h_{j1}-h_{j2})}{h_u}\right) + \\ n\left(\pi D_f L_f + \pi\left(\frac{D_f^2}{4}\right)\right) - n\pi\left(\frac{D_f^2}{4}\right) \end{array} \right)\sigma \tag{9}$$

The remaining portion of the jet has a contribution to surface energy given by

$$s \cdot e_{j2} = A_{j2}\sigma = \pi D_j h_{j2}\sigma \tag{10}$$

Since our focus is on finding the maximum umbrella diameter, $D_u$, the velocity of the jet, $v_{j2}$ at that instant is found to be approximately zero. Therefore

$$k \cdot e_{j2} = 0 \tag{11}$$

Finally, the viscous dissipation work of the droplet as it spreads over the rising jet is expressed by [34]

$$v_{diss} = \int_0^{t_{spread}} \int_0^{V_{dis}} \Phi \, dV dt$$

(12)

Here, $\Phi = \pi_{ji} d_{ij}$ is the viscous dissipation function, where $\pi_{ji}$ is the shear component of deviatoric stress, and $d_{ij}$ is the rate of angular deformation of the liquid. $V_{dis}$ is the volume over which the viscous dissipation occurs over a time $t_{spread}$.

On expanding

$$\Phi = \frac{\mu}{2}\left(\frac{\partial u_i}{\partial x_j} + \frac{\partial u_j}{\partial x_i}\right)\left(\frac{\partial u_i}{\partial x_j} + \frac{\partial u_j}{\partial x_i}\right)$$

(13)

Now, Eqn. (13) can be simplified by scaling laws [34], to get

$$\Phi \approx \mu \left(\frac{U}{\delta}\right)^2$$

(14)

Where, $\mu$ is the liquid viscosity and $\delta$ is the boundary layer thickness.

Now the boundary layer thickness $\delta$ for this interaction can be estimated from the similarity solution for axisymmetric stagnation point flow [35] as

$$\delta = \frac{2D_d}{\sqrt{Re_d}}$$

(15)

Where $Re_d = \frac{\rho U D_d}{\mu}$, is the Reynolds number. Further, we can represent the dissipation volume as,

$$V_{dis} = A_u \delta \qquad (16)$$

Note that for high viscous fluid, the area over which droplet spreads will be reduced by the presence of air film. This has to be taken care of by using an intermediate area $A_i$, where $A_b < A_i < A_u$.

And time required for spreading can be scaled as,

$$t_{spread} \sim \frac{D_d}{U} \tag{17}$$

Therefore, substituting Eqns. (14-17) in Eqn. (12), we finally have the dissipation work as

$$v_{diss} = \mu\left(\frac{U}{\delta}\right)A_u D_d = \frac{\mu U}{2}A_u\sqrt{Re_d} \tag{18}$$

In terms of non-dimensional number, Eqn. (18) can be further reduced to

$$v_{diss} = \frac{\sigma A_u Ca_d{}^2 Re_d{}^{3/2}}{2We_d} \tag{19}$$

Now that we have defined all the terms in the energy balance, equating Eqn. (1) and (3) we have

$$k \cdot e_d + s \cdot e_d + k \cdot e_{j1} + s \cdot e_{j1} = s \cdot e_u + s \cdot e_{j2} + k \cdot e_{j2} + v_{diss} \tag{20}$$

$$\frac{\rho}{2}\left(\frac{\pi D_d{}^3}{6}\right)U^2 + \pi D_d{}^2\sigma + \frac{\rho}{2}\pi\left(\frac{D_j{}^2}{4}\right)h_{j1}v_{j1}{}^2 + \pi D_j h_{j1}\,\sigma + \frac{\pi}{4}D_j{}^2\sigma = \left(2\pi D_u h_u - \pi\left(\frac{D_j{}^2}{4}\right) + \right.$$

$$\left(\frac{\frac{\pi D_d{}^3}{6} + \pi\left(\frac{D_j{}^2}{4}\right)(h_{j1}-h_{j2})}{h_u}\right) + n\left(\pi D_f L_f + \pi\left(\frac{D_f{}^2}{4}\right)\right) - n\pi\left(\frac{D_f{}^2}{4}\right)\right)\sigma + \pi D_j h_{j2}\sigma + 0 +$$

$$\frac{\sigma \pi D_u h_u Ca_d{}^2 Re_d{}^{3/2}}{2We_d} \tag{21}$$

Further to describe the jet dynamics, we define an additional set of non-dimensional numbers, $We_{j1} = \frac{\rho v_{j1}{}^2 D_j}{\sigma}$, and $Re_{j1} = \frac{\rho v_{j1} D_j}{\mu}$.

$$\frac{1}{12}D_d{}^2 We_d + D_d{}^2 + \frac{1}{8}D_j h_{j1} We_{j1} + D_j h_{j1} + \frac{D_j{}^2}{4} = 2D_u h_u - \left(\frac{D_j{}^2}{4}\right) + \left(\frac{\frac{D_d{}^3}{6} + \left(\frac{D_j{}^2}{4}\right)(h_{j1}-h_{j2})}{h_u}\right) +$$

$$n\left(D_f L_f + \frac{D_f{}^2}{4}\right) - n\left(\frac{D_f{}^2}{4}\right) + D_j h_{j2} +$$

$$\frac{D_u h_u Ca_d{}^2 Re_d{}^{3/2}}{2We_d} \tag{22}$$

We further simplify Eqn. (22) to obtain the form

$$D_u{}^* = \frac{\left[1+\frac{We_d}{12}\right]\frac{1}{h_{ud}{}^*}+\frac{D_j{}^*}{h_u{}^*}\left[1+\frac{We_{j1}}{8}-h_j{}^*\right]+D_j{}^{*2}-nD_f{}^*L^*+\frac{D_j{}^*}{4h_{uj}{}^*}-\frac{1}{6h_{ud}{}^{*2}}-\frac{D_j{}^*\left(1-h_j{}^*\right)}{4h_{uj}{}^*h_u{}^*}}{2+\frac{Ca_d{}^2Re_d{}^{\frac{3}{2}}}{2We_d}} \tag{23}$$

Where, $D_u{}^* = D_{max}{}^* = \frac{D_u}{D_d}$, $D_j{}^* = \frac{D_j}{D_d}$, $D_f{}^* = \frac{D_f}{D_d}$, $h_{ud}{}^* = \frac{h_u}{D_d}$, $h_u{}^* = \frac{h_u}{h_{j1}}$, $h_j{}^* = \frac{h_{j2}}{h_{j1}}$, $h_{uj}{}^* = \frac{h_u}{D_j}$, and $L^* = \frac{L_f}{h_u}$.

Eqn. (23) gives us the non-dimensional maximum umbrella diameter, $D_u{}^*$, in terms of the aforementioned non-dimensional numbers. Also, the contribution from $D_j{}^{*2}$ term is less than 0.5%, and can be neglected for the ease of calculations. For cases such as low-ejection heights or high viscosity liquid,where no filaments were formed, $n$ can be taken as zero.

Now in order to validate this model, we measured the experimental value of $D_u$ and other required experimental inputs using ImageJ. We substituted these measured inputs in the model and predicted the maximum umbrella diameter. The results of the validation are shown from Fig. 16 to Fig. 18, for various test liquids and different droplet ejection heights. The error bars are placed with $\pm7.5\%$ from the experimental value. In Fig. 17c, the data was collected for only three droplet ejection heights, 10 cm, 20 cm, and 30 cm. Further, it can also be observed that there is a significant variation in the theoretical results from that experimentally obtained for Water 20X V, as seen in Fig. 18c. This is because for higher than 60% glycerol there is a drop in surface tension from $\sigma \approx 72$ mN/m to $\sigma \approx 65$ mN/m [27], which was not taken into account as we assumed its surface tension to remain unchanged and same as water.

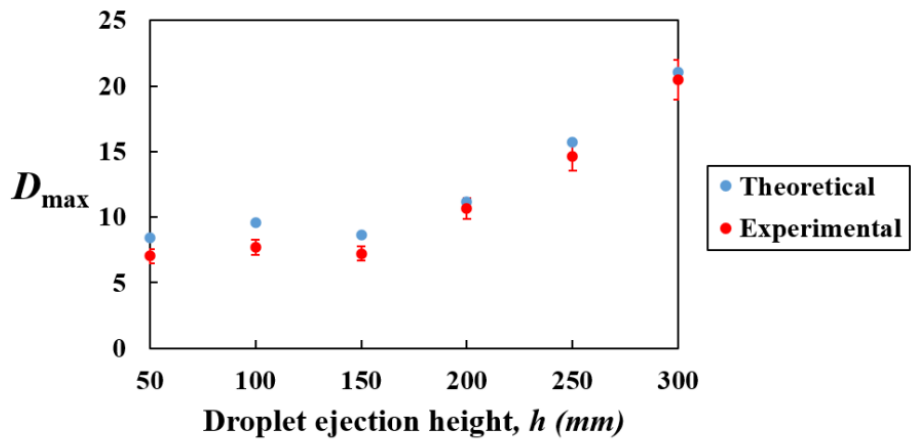

**Fig. 16:** Comparison between theoretical and experimental values of maximum splash diameter for various ejection heights for water.

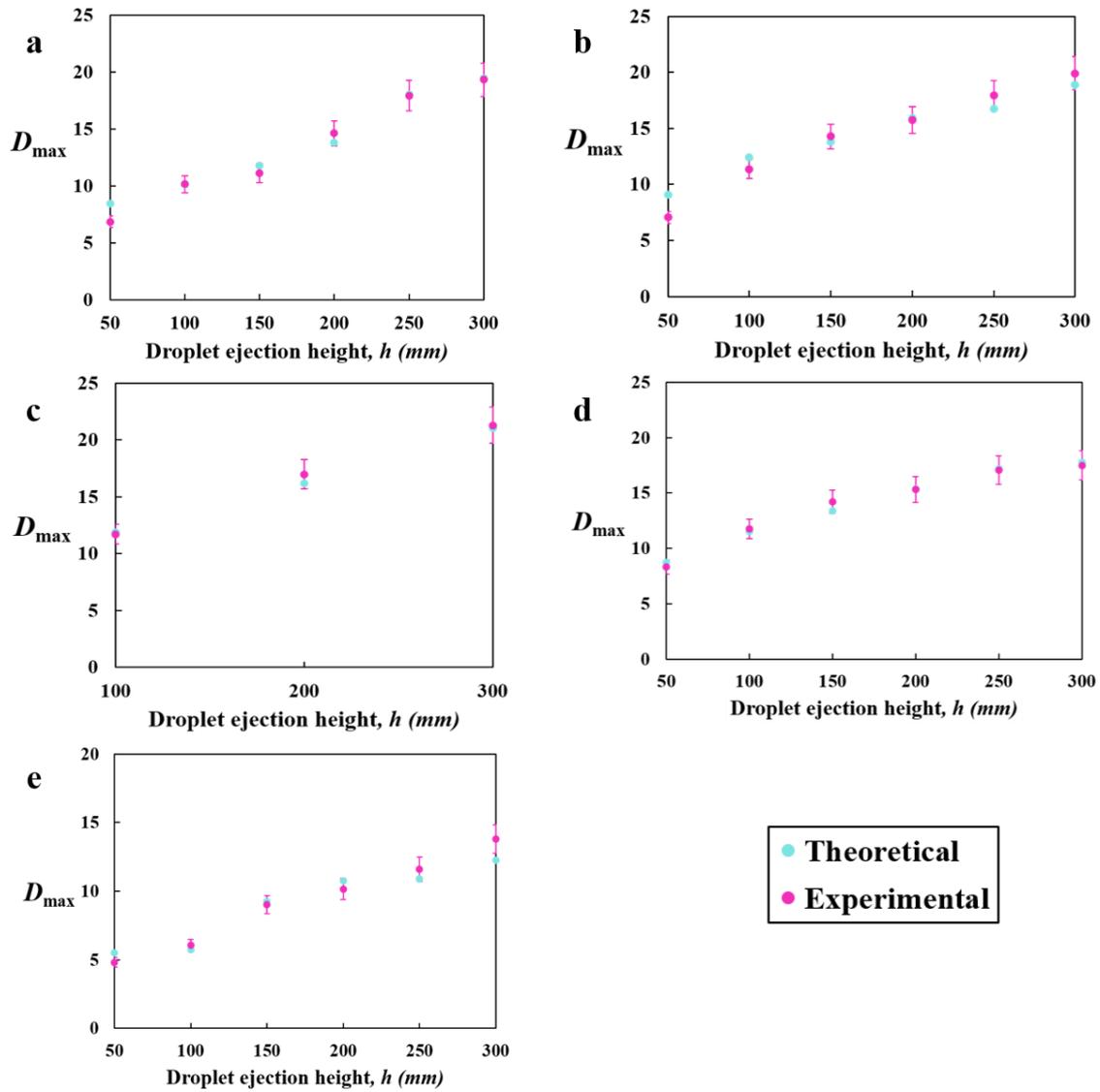

**Fig. 17:** Comparison between theoretical and experimental values of maximum splash diameter for various ejection heights for varying surface tension: a) Water 0.92X ST, b) Water 0.88X ST, c) Water 0.8X ST, d) Water 0.6X ST, e) Water 0.5X ST.

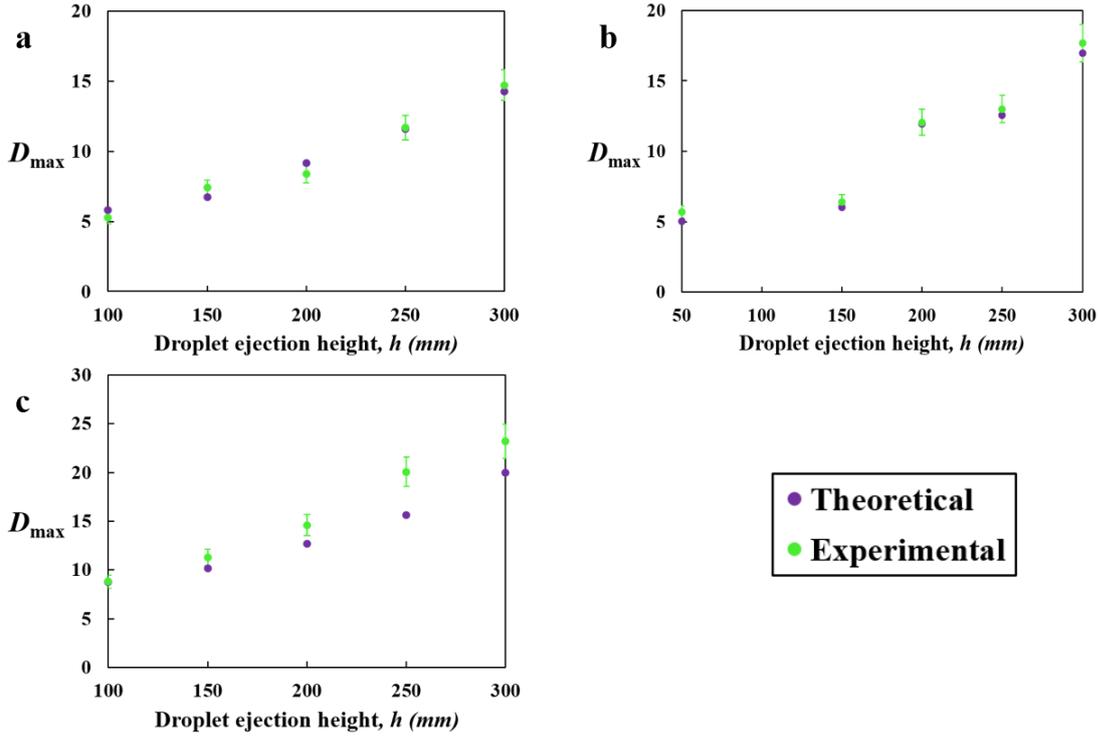

**Fig. 18:** Comparison between theoretical and experimental values of maximum splash diameter for various ejection heights for varying viscosity: a) Water 5X V, b) Water 10X V, c) Water 20X V.

## 4. Conclusions

In the present study on droplet-on-pool impact experiments, the influence of droplet ejection height, surface tension, and viscosity effects on the lamellar structures formed via second-incoming droplet collision with a rising jet/ satellite droplet has been broadly investigated. As the ejection height increases (and thereby the impact velocity of the droplet), primary droplet generates higher-energy jets which then collide with a second droplet to obtain umbrella structures with high splash diameters. This is due to inertial forces from the droplet that push the liquid towards the periphery of the umbrella formed. The collision of the second droplet can be either along the principal axis of the jet/satellite droplet or at an offset to the axis, thereby resulting in two additional configurations: Head-on, Off-set.

Whether the second droplet interacts with a rising jet or a satellite droplet is strongly influenced by the flowrate of the syringe pump. At high flowrate only rising jets are observed, while satellite droplets are observed only when the flowrate is low enough. In the configurations of head-on and off-set, there are certain nuances which set them apart. In effect, for all other conditions remaining unchanged, both should yield approximately the same lamellar structure diameter. This is because in off-set, despite the truncation at one end, the slip at the other end (lower viscous dissipation) compensates for it.

Surface tension tries to minimize the surface energy of a liquid. Hence in droplet collisions, higher degree of spreading is observed with reduction in surface tension. Observation of filaments along the rim of the umbrella/lamella and its subsequent breakup is due to RP instability triggered by surface tension. Therefore, such filaments are a frequent part in surface tension studies.

It is observed that high-viscous liquids undergo a higher degree of spreading. This is due to the entrapment of air film between the incoming second droplet and the rising jet/satellite droplet. As a result, the dissipation occurs over a much smaller area, resulting in wider-diameter umbrellas. Most of the collisions are observed to not produce any filaments in high-viscosity liquids due to a higher degree of order between the molecules.

In addition, a theoretical model based on energy conservation was developed for the head-on collision between a second ejected droplet and rising jet. The model was validated against experimental data (both surface tension and viscosity cases) and was found to be in good agreement.

**Availability of data:** All data relevant to this research are available in this article.

**Conflict of interests:** The authors do not have any conflicts of interest with any individual or agency with respect to this research.

**Acknowledgements:** PD thanks the Department of Mechanical Engineering, IIT Kharagpur for funding this research.